\documentstyle[12pt]{article}

\title{Diffraction in time in terms of Wigner distributions and
tomographic  probabilities
} 

\author{Vladimir Man'ko,$^{a\, \dagger}${\thanks{$^\dagger$Pemanent
address:  P. N. Lebedev Physical Institute,  Leninskii Prospekt 53, 117924
Moscow,  Russian Federation, 
e-mail: manko@na.infn.it ~~~manko@sci.lebedev.ru }  }  
\ Marcos Moshinsky,$^{b\, *}$ \thanks{$^*$Member of 
El Colegio  Nacional} \ Anju Sharma$^{b}$
}   

\date{$^{a}$ Instituto de Ciencias Nucleares-UNAM.\\ 
Apdo. Postal 70-543, 04510 M\'exico, D. F. M\'exico\\
$^{b}$ Instituto de F\'isica-UNAM.\\
Apartado Postal 20-364, 01000 M\'exico, D. F. M\'exico\\
}

\begin{document}

\maketitle

\vfill\eject

\begin{abstract}
Long ago appeared a discussion in quantum mechanics of the problem
of opening a completely absorbing shutter on which were impinging   a
stream of particles of definite velocity. The solution of the problem
was obtained in a form entirely analogous to the optical one of
diffraction by a straight edge. The argument of the Fresnel integrals
was though time dependent and thus the first part in the title of
this article. In section 1 we briefly review the original formulation
of the problem of diffraction in time. In section 2 and 3 we
reformulate respectively this problem in Wigner distributions and
tomographical probabilities. In the former case the probability in
phase space is very simple but, as it takes positive and negative
values, the interpretation is ambiguous, but it gives a classical
limit that agrees entirely with our intuition. In the latter case
we can start with our initial conditions in a given reference frame
but obtain our final solution in an arbitrary frame of reference.
\end{abstract}

\vfill\eject

\section{Introduction}

Long ago~\cite{mo1} one of us (M.M.) discussed in quantum mechanics the
problem of opening at time $t=0$ a completely absorbing shuttler
situated at $x=0$, on which were impinging a stream of particles of
definite velocity. In units in which $\hbar$ and the mass $m$ of the
particles are unity, the problem reduces to finding a wave function
that satisfies the free one dimensional time dependent Schroedinger
equation i.e.,
\begin{equation}
i\frac{\partial \psi (x,t)}{\partial t} = -
%\ahalf 
\frac {1}{2}\frac{\partial^2
\psi (x,t)}{\partial x^2}  ,
\label{eq:in1} 
\end{equation}
with the initial condition
\begin{equation} 
\psi (x,0) = \exp (ikx) \theta (-x),
\label{eq:in2} 
\end{equation}
where $\theta(x)$ is the step function
\begin{equation} 
\theta (x) = \cases{ 1 \quad $if$ \quad x>0\cr 0 \quad $if$\quad x<0},
\label{eq:in3} 
\end{equation}

The solution of this problem was given in reference 1 and later
Nussensveig~\cite{nv} gave it the name $M(x,k,t)$ and it can be
expressed as[1,2,3]  %~\cite{mo1,nv,mo2}
\begin{eqnarray}
&&M(x,k,t) = 
%\ahalf 
\frac {1}{2}
\exp \bigg[ i (kx-
%\ahalf 
\frac {1}{2}
k^2t\bigg] erfc (e^{-i
\pi/4 }w)\nonumber \\
&=& e^{-i\pi/4} \exp \bigg[ i (kx-
%\ahalf 
\frac {1}{2}
k^2t)\bigg] \frac{1}{\sqrt2}
\bigg\{\bigg[ 
%\ahalf 
\frac{1}{2}- C(w)\bigg] + i \bigg[ 
%\ahalf 
\frac {1}{2}
- S(w)\bigg]\bigg\} ,
\label{eq:in4} 
\end{eqnarray}
where
\begin{equation} 
w = \frac{x-kt}{\sqrt{2t}},
\label{eq:in5} 
\end{equation}
and the error integral is
\begin{equation}
	erfc (z) =\frac{2}{\sqrt{\pi}} \int^\infty_z e^{-y^2} dy ,
	\label{eq:in6}
	\end{equation}
while the Fresnel integrals are defined by
\begin{equation} 
C(w) = \sqrt{\frac{2}{\pi}} \int^w_0 \cos y^2 dy, S(w) =
\sqrt{\frac{2}{\pi}} \int^w_0 \sin y^2 dy .
\label{eq:in7}
\end{equation}

Although we have assumed $k$ real, as in the units we use it is the
velocity or momentum of the impinging particles, all the above
expressions remains valid for complex $k$ so long as $Im\, k<0$. In that
case we have the alternative representation~\cite{nv,mo2}
\begin{equation} 
M(x,k,t) =\frac{i}{2\pi} \int^\infty_{-\infty} \frac{\exp
p[i(\kappa x-
%\ahalf 
\frac {1}{2}
\kappa^2t)]}{\kappa-k} d\kappa ,
\label{eq:in8} 
\end{equation}
which follows from the fact that both sides are solutions of
(\ref{eq:in1}) satisfying the initial condition (\ref{eq:in2}).
The Green function for the one dimensional free particle Schroedinger
equation has the form                                    
\begin{equation}
U (x-x',t) = \frac{\exp [i(x-x')^2/2t]}{\sqrt{2\pi it} } ,
 \label{eq:in9}
\end{equation}
as it satisfies Eq. (\ref{eq:in1}) for any $t>0$, but when $t=0$ it
becomes the delta function $\delta(x'-x)$. As our initial condition
is (\ref{eq:in2}) it is clear[1,2] that our function $M(x,k,t)$ can
also be written as
\begin{equation}
M(x,t) = \int^0_{-\infty} U (x-x',t) \exp (ikx') dx'  .
\label{eq:in10}
\end{equation}
The expression  
\begin{equation} 
|M(x,k,t)|^2 ,
\label{eq:in11} 
\end{equation}
gives the probability density of finding the particle at the point
$x$ at time $t$ when initially it was on the left side of the shutter
i.e., with $x<0$ and had a momentum $k$. From (\ref{eq:in4}) we see that
\begin{equation} 
|M(x,k,t)|^2 = 
%\ahalf 
\frac {1}{2}
\bigg\{ \bigg[ 
%\ahalf 
\frac {1}{2}- C(w)\bigg]^2 + \bigg[
%\ahalf 
\frac {1}{2}
- S(w)\bigg]^2 \bigg\} ,
\label{eq:in12} 
\end{equation}
and it is identical to the expression[4] for the intensity of light
in the Fresnel diffraction by a straight edge. The variable $w$ has
though a very different meaning from the optical problem as it is now
a function of time given by (\ref{eq:in5}). Thus the original
paper~\cite{mo1} was given the name ``Diffraction in time''.

All what we said above has been well known for a very long time, and
had many applications among which we wish to mention those related to
the time-energy uncertainty relations~\cite{mo3} and decay
problems~\cite{ga}. 

The reason that we return to this subject is that now we wish to see its
behavior when formulated in terms of Wigner distributions
functions~\cite{wig} and also in relation with the tomographic
probability developed recently by one of us (V.M.) and his
collaborators~\cite{man0}.

\section{Diffraction in time in Wigner distributions space}

\setcounter{equation}{0}

Normally quantum mechanics is discussed in configuration space or, in
some cases, in momentum space, but not in both together.
Wigner~\cite{wig} found that this limitation interfered with the
application of quantum mechanics to the statistical physics where the
description is usually given in phase space. Thus he introduced his
concept of Wigner distributions which allow us to discuss some 
features of quantum mechanics in phase space.

Our objective will be to formulate the diffraction in time problem,
discussed  in the previous section, in terms of Wigner distribution
functions. In this way we can visualize the phenomena in phase space
and more easily determine its classical limit, and compare it with
our intuitive understanding of the behavior of a beam of particles of
definite momentum impinging on a shutter, when the latter is opened.

In units in which $\hbar$ and the mass $m$ of the particle are unity, and
where the configuration space wave function is denoted by $\psi(x)$,
and the momentum by $p$, the Wigner distribution function is defined
as~\cite{wig}  
\begin{equation}
W (x,p) \equiv (\frac1\pi) \int^\infty_{-\infty} \psi^* (x+y) \psi (x-y)
\exp (2ipy) dy ,
\label{eq:di1} 
\end{equation}
which has the  obvious property that
\begin{equation} 
\int^\infty_{-\infty} W (x,p) dp = |\psi (x) |^2,
\label{eq:di2}
\end{equation}
where the right hand side is the probability density at the point
$x$, while an integration with respect $x$ gives us the usual
probability density~\cite{wig} at the momentum value $p$.

If we now wish to discuss the diffraction in time problem in terms of
Wigner distributions we have to replace in (\ref{eq:di1}) $\psi(x)$
by $M(x,k,t)$ of (\ref{eq:in4}).

While for our analysis $k$ is real, we shall assume for the moment
that $k$ is complex with a small negative imaginary part. In this way
we can use the expression (\ref{eq:in8}) for $M(x,k,t)$ and
substituting in (\ref{eq:di1}) we get
\begin{eqnarray}
&&\quad\quad \qquad \qquad   W(x,p;k,t) = \nonumber\\  &&\frac{1}{4\pi^3}
\int^\infty_{-\infty}\int^\infty_{-\infty}\int^\infty_{-\infty} 
\frac{\exp \bigg\{ -i[\kappa(x+y)-     
%\ahalf
\frac {1}{2}
\kappa^2t   ]\bigg\} \exp\bigg\{i
[ \kappa' (x-y)-
%\ahalf
\frac {1}{2}
\kappa'^2t]\bigg\} } {(\kappa-k^*) (\kappa'-k)}\nonumber\\
&&\times e^{i2py} d\kappa d\kappa'dy, 
 \label{eq:di3} 
\end{eqnarray}
where we now added the momentum $k$ and time $t$ to our Wigner
function on the left hand side, as these parameters also appear in the
$M(x,k,t)$. We  also indicate by $k^*$ the complex conjugate of $k$.

The evaluation of the triple integral (\ref{eq:di3}) is done in
Appendix A and it leads to the simple result  
\begin{equation}
W(x,p;k,t) = \frac{1}{2\pi(k-p)} \sin \bigg\{ 2 (pt-x) (k-p)\bigg\}
\theta (pt-x),
\label{eq:di4} 
\end{equation}
where $\theta$ is the step function defined in  (\ref{eq:in3}).
Because of the presence of the sine function in  (\ref{eq:di4}) we see
that the Wigner distribution for the diffraction in time problem
oscillates between positive and negative values, where the physical
significance of the latter is not clear. On the other hand, the
presence of $\theta$ indicates that the probability density in phase
space vanishes when $x>pt$. As in our units $\hbar=m=1$, the momentum
$p$ is the same as the velocity, and this result is intuitively expected
as the particles in our beam with momentum $p$ could not yet have
reached the point $x$.

What is particularly interesting to us is the classical limit
$W(x,p;k,t)$ which is achieved when the Planck constant $\hbar\to 0$. We
have then to abandon units in which $\hbar$ and $m$ were taken as 1
and instead use standard cgs ones. The modifications in the form of
Eq.(\ref{eq:di3}) are trivial and the resulting distribution function
now has the form
\begin{equation}
W(x,p;k,t) = 
%\ahalf 
\frac {1}{2}
\frac{\sin [g(k-p)]}{\pi (k-p)} \theta
\bigg(\frac{pt}{m} -x\bigg),
 \label{eq:di5}
\end{equation}
where           
\begin{equation}
g \equiv \frac{2}{\hbar} \bigg(\frac{pt}{m} -x\bigg).
\label{eq:di6}
\end{equation}

If we take the limit $\hbar \to 0$ then $g\to +\infty$ as the step
function takes the value 1 only if $(pt/m)-x>0$. We can then use one
of the definitions of the $\delta$ function~\cite{gh}
\begin{equation}
\delta (k-p) = \lim_{g\to\infty} \frac{\sin[g(k-p)]}{\pi [k-p]} ,
\label{eq:di7}
\end{equation}
to write the classical limit of our distribution function as
\begin{equation}
W_{cl}(x,p;k,t) = 
%\ahalf 
\frac {1}{2}
\delta (k-p) \theta (\frac{kt}{m} -x) ,
\label{eq:di8} 
\end{equation}
where we  used the presence of the $\delta(k-p)$ in (\ref{eq:di8}) to
replace in the step function the $p$ by $k$.

We now see that our classical limit is what we expect as the only
value possible for the momentum of the particle is $p=k$ and besides
this value is taken only when $x<(kt/m)$, as for $x>(kt/m)$ the
particles would not have yet arrived at the point $x$. Thus the
classical limit of the Wigner distribution function for our
diffraction in time problem confirms our intuition.

\section{Diffraction in time in terms of the tomographic
probabilities}

\setcounter{equation}{0}

In ordinary quantum mechanics the essential concept is the wave
function which in configuration space is denoted by $\psi(x)$. From
this concept one derives the probability density $|\psi(x)|^2$ of
finding the particle at point $x$ and also, through appropriate
transforms of $\psi(x)$, the probabilities for given values of any
other observables.

Recently a change of emphasis has been proposed in which the central
concept is the probability itself, but defining it in a tomographic
way~\cite{man0,man,man1}. 
This allows us to analyze {\it thru a single concept} the
probability either in configuration or momentum space as well as for
variables that are linear combinations of both. The tomographic
probability density~\cite{man0,man1} was given in terms of the Wigner
distribution through the transform 
\begin{equation}
w(X,\mu,\nu) \equiv \frac{1}{2\pi} \int^\infty_{-\infty}
\int^\infty_{-\infty} \int^\infty_{-\infty}  W(x,p)
e^{-iz(X-x\mu-p\nu)} dz dxdp,
\label{eq:te1} 
\end{equation}
where $X$ is the position considered in an ensemble of references
frames~\cite{man0,man1} which are rotated and scaled with respect to the
initial ones through the parameters $\mu,\nu$. As an example we have
that when $\mu=1, \nu=0$ the $X$ corresponds to the normal position
coordinate, but when $\mu=0, \nu=1$, it is related with the momentum
observable. 

In (\ref{eq:te1}) the $W(x,p)$ is the Wigner function defined in
(\ref{eq:di1}) and substituting it in (\ref{eq:te1}) the tomographic
probability density $w(X,\mu,\nu)$ is given in term of the
configuration wave function $\psi(x)$ by  
\begin{equation}
w(X,\mu,\nu) = \frac{1}{2\pi^2}
\int^\infty_{-\infty}\int^\infty_{-\infty}\int^\infty_{-\infty} 
\int^\infty_{-\infty} 
\psi (x-y) \psi^* (x+y) e^{i2py} e^{-iz(X-x\mu-p\nu)}
dzdxdpdy 
\label{eq:te2} 
\end{equation}

The integration with respect to $p$ gives us the expression 
\begin{equation}
\int^\infty_{-\infty} \; dpe^{ip(2y+z\nu)} =\pi \delta
\bigg(y+\frac{z\nu}{2}\bigg) ,
\label{eq:te3} 
\end{equation}
and substituting it in (\ref{eq:te2}), and carrying out the
integration with respect to $y$, we obtain
\begin{equation}
w(X,\mu,\nu) = \frac{1}{2\pi} \int^\infty_{-\infty}
\int^\infty_{-\infty} \psi (x+\frac{z\nu}{2}) \psi^*
(x-\frac{z\nu}{2}) e^{-iz(X-\mu x)} dzdx
\label{eq:te4} 
\end{equation}

Introducing now the variables             
\begin{equation}
u=x+\frac{z\nu}{2} , \qquad r= x-\frac{z\nu}{2},
\label{eq:te5}
\end{equation}
we see the volume element $dzdx$ in (\ref{eq:te4}) becomes
$drdu/|\nu|$, so in terms of $u, r, w(X,\mu,\nu)$ becomes~\cite{man}
\begin{eqnarray}
&&w(X,\mu,\nu) = \frac{1}{2\pi|\nu|} 
\int^\infty_{-\infty} \int^\infty_{-\infty}
\psi (u)\psi^*(r) \exp \bigg\{-i \frac{u-r}{\nu} \bigg[ x-\mu
\bigg(\frac{r+u}{2}\bigg)\bigg]\bigg\} 
drdu \nonumber\\
&&\quad \quad = \frac{1}{2\pi|\nu|} \bigg| \chi (X,\mu,\nu)\bigg|^2 ,
\label{eq:te6} 
\end{eqnarray}
where           
\begin{equation} 
\chi(X,\mu,\nu) = \int^\infty_{-\infty} \psi (u) \exp [i \bigg(
\frac{\mu}{2\nu} u^2 - u \frac{X}{\nu}\bigg) \bigg] du .
\label{eq:te7}
\end{equation}

Thus, contrary to the Wigner distribution function, the tomographic
probability density is always positive definite.

We now turn to the problem of diffraction in time which means
replacing $u$ by $x$ in (\ref{eq:te6}) and then $\psi(x)$ by
$M(x,k,t)$ given in terms of its expression (\ref{eq:in10})
containing the Green function of the free particle motion. The expression
$\chi(x,\mu,\nu)$ takes then the form 
\begin{equation}
\chi (X,\mu,\nu) = \int^\infty_{-\infty} \int^0_{-\infty} \frac{\exp
[i(x-x')^2/2t]}{\sqrt{2\pi(it)}} e^{ikx'} e^{-iX\frac{x}{\nu} }
e^{\frac{i\mu x^2}{2\nu} } dx dx'
\label{eq:te8}
\end{equation}

This integral is evaluated in a straight forward but laborious way in
Appendix B, where its value is given. As we are only interested in its
absolute value squared multiplied by $(2\pi|\nu|)^{-1}$ which, from
(\ref{eq:te6}) gives us the tomographical  probability density, we
see that it becomes
\begin{equation}
w(X,\mu,\nu) = \frac{1}{2|\mu|} \bigg\{ \bigg[ 
%\ahalf 
\frac {1}{2}
+
C(\rho)\bigg]^2 + \bigg[ 
%\ahalf 
\frac {1}{2}
+ S(\rho)\bigg]^2\bigg\} ,
 \label{eq:te9} 
\end{equation}
where           
\begin{equation}
\rho=\frac{k(\mu t +\nu)-X}{\sqrt{2\mu(\mu t+\nu)}}
\label{eq:te10},
\end{equation}
and $C,S$ are the Fresnel integrals defined in Eq.(\ref{eq:in7}).

We proceed now to discuss the meaning of the tomographical
probability density given in (\ref{eq:te8}). We mentioned above that
$\mu,\nu$  represent a rotation and scaling of an ensemble of
reference frames in phase space with respect to the original one.
Thus we can express them as     
\begin{equation} 
\mu = e^\tau \cos \theta, \qquad \nu = e^{-\tau} \sin \theta ,
\label{eq:te11} 
\end{equation}
with $\tau,\theta$ in the intervals $-\infty\leq \tau \leq \infty,
0\leq \theta \leq 2 \pi$. These expressions of $\mu, \nu$ imply that
our new coordinate and momenta, which we designate by capital $X,P$,
are given in terms of the original ones, which we denote by lower
case letters $x,p$, through the relation~\cite{mo4}
\begin{equation}
\pmatrix {X\cr P} = \pmatrix { e^\tau \cos \theta &e^{-\tau} \sin
\theta\cr -e^\tau \sin \theta &e^{-\tau} \cos\theta} \pmatrix {x\cr
p} ,
\label{eq:te12}
\end{equation}
which is a linear canonical transformation as the determinant of the
matrix is 1. The $w(X,\mu,\nu)$ of (\ref{eq:te9}), with $\rho$ given by
(\ref{eq:te10}), gives then the probability density for the
diffraction in time problem in the new configuration coordinate $X$
defined in (\ref{eq:te12}).

If we want to return to our original configuration space we see from
(\ref{eq:te12}) that we must take there $\tau=\theta=0$ which implies
$X=x$ and $\mu=1, \nu=0$.

In that case $\rho$ of (\ref{eq:te10}) becomes
\begin{equation} 
\rho= \frac{kt -x}{\sqrt{2t}} = -w ,
\label{eq:te13} 
\end{equation}
where $w$ was defined in (\ref{eq:in5}). As the Fresnel integrals are
odd functions of the argument we have from (\ref{eq:te13}) that
\begin{equation}
C(\rho) = - C(w), \qquad S(\rho) = - S(w) ,
\label{eq:te14}
\end{equation}
and thus the particular tomographic density $w(x,1,0)$ becomes                              
\begin{equation}
w(x,1,0) = 
%\ahalf 
\frac {1}{2}\bigg\{ [
%\ahalf 
\frac {1}{2}
- C(w)]^2 + [
%\ahalf 
\frac {1}{2}
- S(w)]^2
\bigg\} ,
\label{eq:te15}
\end{equation}
which is identical to the expression (\ref{eq:in12}), as we should
expect.

Thus we see that the analysis of diffraction  in time phenomena in
terms of the tomographic probabilities, allows us to study the phenomena in
a wide ensemble of reference frames in phase space as indicated in
Eqs.~(9--11).
%(\ref{eq:te8})--(3.11).

This ensemble includes of course the original phase space $(x,p)$ in
which the result is given by (\ref{eq:te14}) agreeing exactly with
the initial analysis of the problem~\cite{mo1}.

\section{Conclusion}

In the present paper the problem of diffraction in time was
visualized from three different viewpoints. The first was the
original one~\cite{mo1}, in which both the initial conditions and the
solution of the problem were analyzed in this same frame of
reference. The solution (\ref{eq:in5}) was given in terms of the Fresnel
integrals, and using Cornu spiral we showed that the usual
diffraction pattern appeared as function of time.

In the second approach we translated our solution to the Wigner
distribution space. The final expression for the probability density
in phase space turned out to be very simple but, unfortunately, it
could take both positive and negative values, which made its
interpretation ambiguous.

Fortunately it was possible to consider its classical limit by taking
 $\hbar \to 0$, and the resulting expression (\ref{eq:di8}) agreed
entirely with our intuitive view, i.e., the probability in phase space
was only different from zero when $p=k$ and $x<(pt/m)$, where all the
observables are in cgs units.

The third approach implied formulating our solution in terms of
tomographic probabilities. The latter have been introduced
recently [8,10,11] %~\cite{man0,man,man1} 
to allow us to express the solutions in any
reference frame that is rotated and scaled with respect to original
one. In effect it implies carrying out a canonical transformation on
the original solution of the diffraction in time problem. Our
tomographic probability solutions (\ref{eq:te9}) is again expressed
in terms of Fresnel integrals but of an argument quite different from
the one appearing (\ref{eq:in5}). If the canonical transformations is
the unit one i.e., $X=x, P=p$, then our tomographic probability reduces
to the solution (\ref{eq:in4}), providing us with a check of the
analysis developed in section 3.

We finally wish to indicate that the diffraction in time phenomena
derived theoretically in 1952 was, in a somewhat changed
form~\cite{sz}, measured experimentally in 1996. Possibly a similar
fate, in a distant future, awaits the reformulation of the phenomena
presented in this paper.

\vfill\eject

\centerline{\large\bf Appendix A: Determination the Wigner function
$W(x,p;k,t)$. }

\vskip20pt

We start with the expression (\ref{eq:di3}) for $W(x,p;k,t)$ and
rewrite it as
$$
 W (x,p;k,t) = \frac{1}{4\pi^3}
\int^\infty_{-\infty}\int^\infty_{-\infty} \bigg\{
\int^\infty_{-\infty} \exp [2ipy \bigg( p-\frac{\kappa
+\kappa'}{2}\bigg) \bigg] dy \bigg\}
$$
$$\times \bigg\{ \frac{\exp [-i(\kappa x-
%\ahalf
\frac {1}{2}
\kappa^2 t
)]}{(\kappa-k^*)} \frac{\exp [i(\kappa' x -  
%\ahalf
\frac {1}{2}
\kappa'^2 t
)]}{(\kappa'-k)}\bigg\} d\kappa d\kappa' .
\eqno(A.1)$$

The first integral obviously gives the $\delta$ function $2\pi\delta
(p-\frac{\kappa+\kappa'}{2})$ and so introducing it in (A.1) and
integrating with respect to $\kappa'$ we obtain
$$
W(x,p;k,t) = - \frac{1}{2\pi^2} \exp [2ip(x-pt)] \int^\infty_{-\infty}
\frac{\exp[-2i\kappa(x-pt)]}{(\kappa-k^*) (\kappa+k-2p)} d\kappa .
\eqno(A.2)$$
As we have that
$$
\frac{1}{(\kappa-k^*) (\kappa+k-2p)} = \frac{1}{(k+k^*-2p)} \bigg[
\frac{1}{(\kappa-k^*)} -\frac{1}{(\kappa+k-2p)}\bigg] ,
\eqno(A.3)
$$
we see that by introducing it in (A.2) we get
$$ W(x,p;k,t) = - \frac{\exp[2ip (x-pt)]}{(2\pi)^2 (k+k^*-2p)}$$
$$\times \bigg\{ \int^\infty_{-\infty} d\kappa \frac{\exp
[-2i\kappa(x-pt)]}{(\kappa-k^*)} - \int^\infty_{-\infty} d\kappa
\frac{\exp  [-2i\kappa (x-pt)]}{(\kappa+k-2p)} \bigg\} . \eqno(A.4)
$$
We now note, as indicated in the text after Eq(\ref{eq:di3}), that we
start by assuming that $k$ has as a small negative imaginary part so
that
$$
k \to k - i\epsilon; \; k^* \to k + i\epsilon; \; -k +2p \to -k+2p
+i\epsilon
\eqno(A.5)$$

Thus the singularity in the integrals in (A.4) is in the upper half
of the $\kappa$ plane. 

We can close the contour in (A.4) by a  large circle in the upper
half of the complex $\kappa$ plane if $x-pt<0$ thus getting the residues
of the integrals at the points $k+i\epsilon, - k+2p+i\epsilon$. On
the other hand if $x-pt>0$ we have to close the contour by a large
circle in the lower half plane and, as the function is analytic
inside the contour, the integral vanishes. Then, passing to the limit
when $\epsilon \to 0$, as required for our problem where $k$ is real,
we get, after carrying some of the multiplications, that
$$ 
W(x,p;k,t)= \frac{\theta (pt-x)}{4\pi i(k-p)} \bigg\{ \exp [2i
(k-p) (x-pt)] - \exp [-2i (k-p) (x-pt)] \bigg\}.
\eqno(A.6)$$
where $\theta$ is the step function (\ref{eq:in3}). As the curly
bracket derived by $2i$ is a sine function we then obtain the
expression (\ref{eq:di4}).

\vfill\eject

\centerline{\large\bf Appendix B: Determination of the tomographic
probability $w(X,\mu,\nu)$ }

\vskip20pt

The tomographic probability is proportional to the absolute square of
$\chi(X,\mu,\nu)$ where the latter is given by (\ref{eq:te7}) and we
rewrite it in the form
$$
w(X,\mu,\nu) = \int^0_{-\infty} \bigg\{ \frac{\exp [ikx' +
i(x'^2/2t)]}{\sqrt{2\pi it}} \int^\infty_{-\infty} \exp [i (ax^2 -bx)]
dx\bigg\} dx' , \eqno(B.1)
$$
where
$$
a\equiv \frac{\mu}{2\nu} + \frac{1}{2t}, \qquad b\equiv 
\frac{X}{\nu} + \frac{x'}{t}, \eqno(B.2) $$

We can rewrite the expressions in the last round bracket in (B.1)
getting
$$
ax^2 - bx = \bigg(\sqrt{a} x - \frac{b}{2\sqrt{a}} \bigg)^2 - \bigg(
\frac{b^2}{4a}\bigg) . \eqno(B.3)
$$
As $b^2/4a$ depends on $x'$ but not on $x$, we first evaluate
the integral
$$
\int^\infty_{-\infty} \exp \bigg[i \bigg(\sqrt{a} x - \frac{b}{2\sqrt{a}}
\bigg)^2\bigg] dx = \sqrt{\frac{\pi}{a} } e^{i\pi/4}, \eqno(B.4)
$$
to obtain
$$
\chi (X, \mu,\nu) = \frac{e^{i\pi/4}}{\sqrt{i (\frac{\mu t}{\nu} +1)}} 
\int^0_{-\infty} \exp [ikx' +i (x'^2/2t)]\exp
\bigg[ \frac{-i(\frac{x'}{t}+\frac{X}{\nu})^2}{2(\frac{\mu}{\nu} +
\frac{1}{t})} \bigg] dx' $$
$$
=\frac{e^{i\pi/4} \exp [\frac{-iX^2}{2\nu^2}
(\frac{\mu}{\nu}+\frac{1}{t})^{-1}] }
{\sqrt{i(\frac{\mu t}{\nu} + 1)}}
\int^0_{-\infty} \exp [i (\alpha x'^2 + \beta x')] 
		dx' , \eqno(B.5)
$$
where
$$
\alpha = \frac{(\mu/\nu)}{2(\frac{\mu t}{\nu} +1)}, \qquad \beta =
\frac{k(\frac{\mu t}{\nu} +1) - \frac{X}{\nu}}{(\frac{\mu t}{\nu} +1)},
\eqno(B.6)
$$
Using again the relation (B.3) we get
$$
\alpha x'^2 + \beta x' = (\sqrt{\alpha} x' +
\frac{\beta}{2\sqrt{\alpha} } )^2 - \frac{\beta^2}{4\alpha},
\eqno(B.7)$$
and as $\beta^2/4\alpha$ is independent of $x'$ we need to consider first the
integral
$$
\int^0_{-\infty}  e^{i(\sqrt{\alpha} x' 
+ \frac{\beta}{2\sqrt{\alpha} })^2} dx' = 
\int^{\frac{\beta}{2\sqrt{\alpha}}}_{-\infty} e^{iy^2}
\frac{dy}{\sqrt{\alpha} } =\int^0_{-\infty} e^{iy^2} 
\frac{dy}{\sqrt{\alpha} } +
\frac{1}{\sqrt{\alpha}} \int^{\frac{\beta}{2\sqrt{\alpha}}}_0 
(\cos y^2 + i \sin y^2 ) dy
$$
$$ = \frac{\sqrt{\pi}}{2\sqrt{\alpha}} \frac{(1+i)}{\sqrt2} +
\frac{1}{\sqrt{\alpha}} \frac{\sqrt{\pi}}{\sqrt2} [ C
\bigg(\frac{\beta}{2\sqrt{\alpha}}) + i S
(\frac{\beta}{2\sqrt{\alpha}})\bigg] \eqno(B.8)
$$
where $C,S$ are the Fresnel integrals of (1.7) and $\alpha,\beta$ are
given by (B.6).

Using (B.7) to introduce (B.8) in (B.5) we obtain 
$$
\chi (X, \mu,\nu) = \frac{\sqrt{\pi} e^{i\pi/4}}{\sqrt{2 (\frac{\mu
t}{\nu} + 1) \alpha}} \exp (-i\beta^2/4\alpha) 
	\exp \bigg\{ -i (X^2/2\nu^2) \bigg(\frac{\mu}{\nu} +
\frac{1}{t}   \bigg)^{-1} \bigg\}
$$
$$
\times \bigg\{ \bigg[ 
%\ahalf 
\frac {1}{2}
+ C (\frac{\beta}{2\sqrt{\alpha}})\bigg]
+ i \bigg[ 
%\ahalf 
\frac {1}{2}
+ S (\frac{\beta}{2\sqrt{\alpha}}) \bigg] \bigg\} .
\eqno(B.9) 
$$

Finally replacing $\alpha,\beta$ by their values (B.6) we get.
$$
\chi (X,\mu,\nu) = \frac{\sqrt{\pi} e^{i\pi/4}}{\sqrt{(\mu/\nu)}}
\exp \bigg[-i (X^2/2\nu^2) \bigg(\frac{\mu}{\nu} 
+ \frac{1}{t}\bigg)^{-1} \bigg]
$$
$$
\exp (-i\rho^2) \bigg\{ \bigg[
%\ahalf 
\frac {1}{2}
+ C(\rho)] + i[ 
%\ahalf 
\frac {1}{2}
+
S(\rho)\bigg] \bigg\},
\eqno(B.10)
$$
where
$$
\rho=\frac{k (\mu t +\nu) - X}{\sqrt{2\mu (\mu t + \nu)}} .
\eqno(B.11)$$

When taking the absolute square value of $\chi (X, \mu,\nu)$ mainly
the curly bracket remains and thus we get Eq.(\ref{eq:te8}) whose
properties are discussed in the main text.

\vfill\eject

\end{document}